\documentclass[conference, 9pt]{IEEEtran}
\IEEEoverridecommandlockouts

\usepackage{cite}
\usepackage{amsmath,amssymb,amsfonts}
\usepackage{algorithmic}
\usepackage{graphicx}
\usepackage{textcomp}
\usepackage{xcolor}

\usepackage{colortbl}

\usepackage{pifont}  
\def\BibTeX{{\rm B\kern-.05em{\sc i\kern-.025em b}\kern-.08em
    T\kern-.1667em\lower.7ex\hbox{E}\kern-.125emX}}
\usepackage[bookmarks=true,breaklinks=true,letterpaper=true,colorlinks,citecolor=blue,linkcolor=blue,urlcolor=blue]{hyperref}
\newcommand{\secref}[1]{Section~\ref{#1}}
\newcommand{\figref}[1]{Figure~\ref{#1}}
\newcommand{\tabref}[1]{Table~\ref{#1}}
\begin{document}

\title{ VEDA: Efficient LLM Generation Through \underline{V}oting-based KV Cache \underline{E}viction and \underline{D}ataflow-flexible \underline{A}ccelerator\\ \vspace{-5pt}}

\author{\IEEEauthorblockN{Zhican Wang$^{1}$, Hongxiang Fan$^{2}$, Haroon Waris$^{3}$, Gang Wang$^{1}$, Zhenyu Li$^{1}$, Jianfei Jiang$^{1}$,  Yanan Sun$^{1}$, Guanghui He$^{1}$$^{*}$}
\IEEEauthorblockA{\textit{$^{1}$State Key Laboratory of Micro/Nano Engineering Science, School of Electronic Information and Electrical Engineering,} \\
\textit{Shanghai Jiao Tong University, Shanghai, China,} \\ \textit{$^{2}$Imperial College London, United Kingdom,} \textit{$^{3}$Institute of Space Technology, Pakistan}\\
\{wang\_zhican,wangganginjstu, ambitious-lzy, jiangjianfei, sunyanan, guanghui.he\}@sjtu.edu.cn}, hongxiangfan@imperial.ac.uk, haroon.waris@ist.edu.pk
\thanks{This work was supported by the National Science Foundation of China under Grant 92464302. ($^*$Corresponding author: Guanghui He).\vspace{-1em} }
}

\maketitle
\begin{abstract}
Large Language Models (LLMs) excel in natural language processing tasks but pose significant computational and memory challenges for edge deployment due to their intensive resource demands. This work addresses the efficiency of LLM inference by algorithm-hardware-dataflow tri-optimizations. We propose a novel \textit{voting-based} KV cache eviction algorithm, balancing hardware efficiency and algorithm accuracy by adaptively identifying unimportant kv vectors. From a dataflow perspective, we introduce a \textit{flexible-product} dataflow and a runtime reconfigurable PE array for matrix-vector multiplication. The proposed approach effectively handles the diverse dimensional requirements and solves the challenges of incrementally varying sequence lengths. Additionally, an \textit{element-serial} scheduling scheme is proposed for nonlinear operations, such as softmax and layer normalization (layernorm). Results demonstrate a substantial reduction in latency, accompanied by a significant decrease in hardware complexity, from $O(N)$ to $O(1)$. The proposed solution is realized in a custom-designed accelerator, \textit{VEDA}, which outperforms existing hardware platforms. This research represents a significant advancement in LLM inference on resource-constrained edge devices, facilitating real-time processing, enhancing data privacy, and enabling model customization.

\end{abstract}

\section{Introduction}
\label{intro}
Large language models (LLMs) \cite{achiam2023gpt,touvron2023llama2openfoundation} have demonstrated remarkable performance across various natural language processing tasks, such as machine translation, text summarization, dialogue systems, and code generation. This potential has led to a growing trend of deploying private LLMs on edge devices (e.g., personal computers, smartphones, robotics, and intelligent vehicles), enabling real-time inference, data privacy, and model customization. However, the impressive performance of LLMs comes at the cost of significant computational effort and memory storage requirements, exacerbating the disparity between the substantial resource demands of LLMs and the limited resources available on edge devices.

The inference process of LLMs can be summarized in two phases: prefilling and generation. \underline{In the prefilling phase}, input tokens are provided to the model in parallel, and the model encodes these tokens to kv vectors and concatenates them to form the KV cache. The primary operation in this phase is general-purpose matrix multiplication (GEMM), which is typically compute-bound. This process is similar to that of encoder-only Transformers, such as BERT and ViT\cite{devlin2019bertpretrainingdeepbidirectional,dosovitskiy2021imageworth16x16words}, with numerous dedicated accelerators being proposed for this task\cite{a3,spatten,lu2021sanger}. \underline{In the generation phase}, tokens are generated sequentially in an auto-regressive manner. Each token attends to the entire KV cache formed in previous steps and extends the KV cache by adding the kv vector of the current token. Since only one token is generated at each step, the primary operation is general-purpose matrix-vector multiplication (GEMV), which is memory-bound. Very few solutions have been proposed to optimize the latter phase. Recent work \cite{yu2022orca} suggests batch scheduling to enable weight sharing across multiple requests, which packages GEMV operations into GEMM for linear layers. While this is effective in cloud environments with multi-batch processing, edge devices typically use single-batch inference. Additionally, although batch scheduling improves performance for linear layers, it has limited impact on the attention process, as each user has a distinct KV cache, requiring the use of GEMV operations. Therefore, to enhance LLM inference efficiency on edge devices, particularly in the generation phase, this work proposes a comprehensive solution encompassing algorithmic improvements, dataflow optimization, and hardware architecture.

\textbf{From the algorithmic perspective,} numerous studies have shown that the attention process in LLM inference is highly sparse, with sparsity levels approaching 95\% \cite{zhang2023h2o}. This means that only a small fraction of the KV cache is actively attended to, suggesting the feasibility of locating and maintaining only the pivotal positional kv vectors while discarding the others. Reducing the size of the KV cache not only decreases the computational effort required for critical modules of LLMs (e.g., the attention process) but also reduces off-chip memory access and memory consumption. However, existing algorithms have several limitations. \textbf{Some are simple to implement but suffer from accuracy loss.} For example, \cite{xiao2023efficient} proposes a sliding window strategy that retains only the earliest and most recent positional KV caches. While simple, it loses accuracy due to the inherent information loss from discarded out-of-window KV caches. \cite{zhang2023h2o} introduces a KV cache eviction strategy based on accumulated attention scores, but this approach has its own limitations, which we will discuss in this paper. \textbf{Other methods offer better accuracy but involve complex operations,} such as real-time top-k engines\cite{spatten}, complicated formula \cite{adnan2024keyformer}, which are inefficient for deployment. To achieve both hardware efficiency and algorithmic accuracy, we propose a novel \textit{voting-based KV cache eviction} algorithm. Our approach is inspired by the voting process in democratic systems, where each token acts as a ``voter" with a unique attention score. This score inherently reflects each token's attention and relation to the kv vectors of prior tokens. Each token ``votes" to determine which kv vector should be evicted, and by synthesizing these votes, we can efficiently remove unimportant kv vectors. This algorithm achieves superior accuracy compared to existing methods while maintaining hardware simplicity.

\textbf{From the dataflow and hardware perspectives,} we identify several challenges and propose corresponding solutions. \textbf{For GEMV operations:} \underline{First, widely ranged dimension size,} the dimension size of matrix-vector multiplications varies significantly across different models and even within the same model. For example, in a $(1, k)\times(k, n) = (1, n)$ multiplication, the size of $k$ can range from as small as 128 to as large as 11008 in Llama $7$B. \underline{Second, dynamic inference process,} in the generation phase, the KV cache increases incrementally with each step, meaning that the dimensions of $k$ or $n$ change dynamically. However, existing GEMV accelerators, such as a multiplier array equipped with an adder tree, are designed with a fixed addition size, which leads to underutilization when $k$ is smaller than the fixed size. For instance, suppose the multiplier array and the adder tree can perform $256-$ input multiplication and addition in parallel, but if the $k$ is less than $256$, the multiplier and adder tree are underutilized. Similarly, if $k$ increases from $256$ to $257$, it increases one epoch of computation from $1$ to $2$, and the hardware is extremely underutilized for the second epoch. \underline{More importantly,} the inability to utilize the sequence variation may make KV cache reduction less helpful to inference speed. \underline{Third, memory access irregularity,} the transpose operation on the K matrix hinders inference due to irregular memory access patterns. Storing these vectors in the correct format offline is impractical because they are generated online rather than as static weights. While dedicated hardware modules could solve this issue, they would sacrifice efficiency. To mitigate these dataflow inefficiencies, we resort to mathematical interpretations for inspirations. Since GEMV can be interpreted as either inner-product or outer-product calculations mathematically, we propose \textit{flexible-product dataflow} with a runtime reconfigurable GEMV PE array. This dataflow, combined with the dedicated PE array, efficiently handles size variations and eliminates the need for transpose operations, resulting in optimized inference.

\textbf{For nonlinear operators,} especially softmax and layernorm, we face several challenges. \underline{First,} these operators introduce data dependencies that prevent parallel computation until all required data is available. \underline{Second,} they occur among GEMV operations, and their latency can block subsequent GEMV computations. \underline{Third,} the exponentiation and division operations required by these operators are more computationally expensive than multiplications and additions, and placing sufficient hardware resources for these operations can reduce latency but introduces significant overhead.  To address these unprecedented challenges, we leverage the \textit{flexible-product} dataflow and further propose a \textit{element-serial scheduling} scheme. This approach not only resolves the latency bottleneck through parallel processing but also reduces the number of high-cost modules from $O(N)$ to $O(1)$ through architectural co-optimization.

Finally, integrating these comprehensive optimizations, we present \textit{VEDA}, a dedicated LLM accelerator. To the best of our knowledge, this is one of the pioneer works to apply algorithm-hardware-dataflow tri-optimizations to accelerate LLM generation. Extensive results demonstrate the efficiency of our solution. In summary, we make the following contributions:
\begin{itemize}
    \item We propose a novel \textit{voting-based KV cache eviction} algorithm for LLM acceleration, achieving both efficiency and accuracy.
    \item We provide an insightful analysis of the challenges posed by GEMV and nonlinear operators in LLMs from both hardware and dataflow perspectives.
    \item We propose a \textit{flexible-product} dataflow with a runtime reconfigurable PE array to address GEMV challenges and introduce \textit{element-serial scheduling} to resolve the nonlinear operator bottleneck.
    \item We design \textit{VEDA}, a dedicated LLM accelerator that efficiently supports both prefilling and decoding.
\end{itemize}

\section{Background}
\begin{figure}[hbt]
\vspace{-15pt}
\centering
\includegraphics[width=80mm]{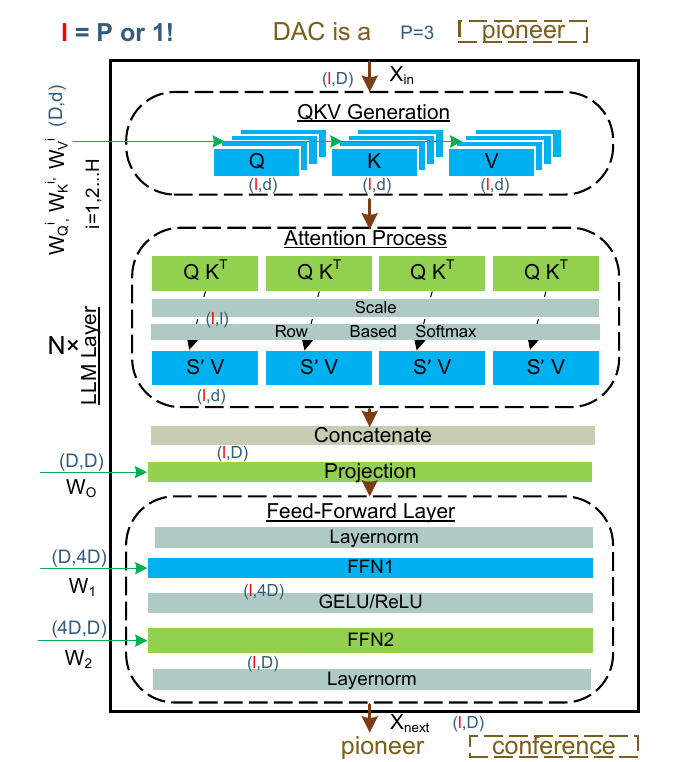}
\vspace{-10pt}
\caption{Detailed LLM structure with sizes denoted. In this paper, matrix-form activations are represented in capital letters(e.g., KV cache), while lowercase letters denote vector-wise activations (e.g., attention score vector $s'$). Green and blue differentiate the dataflows, which collectively  provide the optimal dataflow for LLM, which will be introduced in \secref{sec:nonlinear}}
\label{Fig:LLM}
\vspace{-10pt}
\end{figure}

\figref{Fig:LLM} illustrates the structure of an LLM, with typical dimension sizes annotated. An input sequence, denoted as $X_{in}$, has a length of $l$ tokens, each with a hidden dimension size of $D$, represented as $(l, D)$. The LLM computation consists of four main steps: \underline{Step 1 - QKV Generation}: The input $X_{in}$ is multiplied by $W_{Q}^{i}$, $W_{K}^{i}$, and $W_{V}^{i}$ to generate $Q^{i}$, $K^{i}$, and $V^{i}$ for $i = 1, 2, ..., H$, where $H$ is the number of attention heads. $Q^{i}$, $K^{i}$, and $V^{i}$ all have dimensions of $(l, d)$, with $D = Hd$, where $d$ represents the head dimension. \underline{Step 2 - Attention Process}: Within each head, $Q^{i}$ is multiplied by ${K^{i}}^{T}$ to produce $S^{i}$ of size $(l, l)$. Notice that due to the causality of LLM, each token only relates to the token before it, the upper triangle of $S^{i}$ needs to be set $-\infty$. Each element of $S^{i}$ is scaled to ensure numerical stability, followed by a row-wise softmax operation to generate the attention score matrix $S'^{i}$. $S'^{i}$ is then multiplied by $V^{i}$, resulting in the output matrix $O^{i}$ with dimensions $(l, d)$. \underline{Step 3 - Linear Projection}: The outputs $O^{i}$ from all heads are concatenated into a single matrix of size $(l, D)$, which is then multiplied by $W_o$. \underline{Step 4 - Feed-forward Layer}: In this step, the input is multiplied by $W_1$ of size $(D, 4D)$ in FFN1, producing an intermediate result of size $(l, 4D)$, followed by a non-linear activation function like GELU or ReLU. The output is then multiplied by $W_2$ of size $(4D, D)$ in FFN2, after performing the layernorm (or RMSnorm), it produces the final output $X_{next}$.

\underline{LLM Inference Phase}: LLM inference is divided into a prefilling phase and a generation phase. In the prefilling phase, $l$ corresponds to the prompt length $P$. For example, in \figref{Fig:LLM}, the prompt is ``DAC is a" with $P=3$. A key-value (KV) cache is constructed for each token during this phase, ultimately generating the output token ``pioneer", which marks the end of the prefilling phase.
During the generation phase, $l$ is always one, and activations are reduced to vectors. In the example, ``pioneer" serves as input to the LLM, engaging in an attention process with previous KV caches and extending the KV cache with the current key-value vector. The output ``conference" is then generated as the next token.

\section{Voting-based KV Cache Eviction}
\textbf{Pivotal and Unimportant kv Vectors Identification.} The key challenge in effective KV cache eviction is that we cannot preemptively determine which KV tensors are pivotal, and which are unimportant at each new token generation step. Fortunately, some studies suggest that many pivotal kv vectors are likely to remain pivotal, and vice versa \cite{liu2024scissorhands}. However, defining the importance of kv vectors is non-trivial, and accurately predicting the preferences of newly generated tokens remains challenging. Some research indicates that the most recent tokens are often pivotal, leading to the use of sliding window attention to maintain a fixed KV cache size. However, this approach suffers from accuracy degradation due to the loss of out-of-window tokens. Another commonly adopted method utilizes accumulated attention scores to define the importance of a kv vector. As illustrated in \figref{Fig:comp} (a), the method accumulates the attention scores vertically across each column, generating an importance vector with a length matching the KV cache. Each element of the importance vector corresponds to the significance of each positional KV. The positional KV with the minimum value in the importance vector is selected for eviction, such as position $6$ in the figure. However, several limitations have been identified:
\begin{figure}[hbt]
\vspace{-15pt}
\centering
\includegraphics[width=88mm]{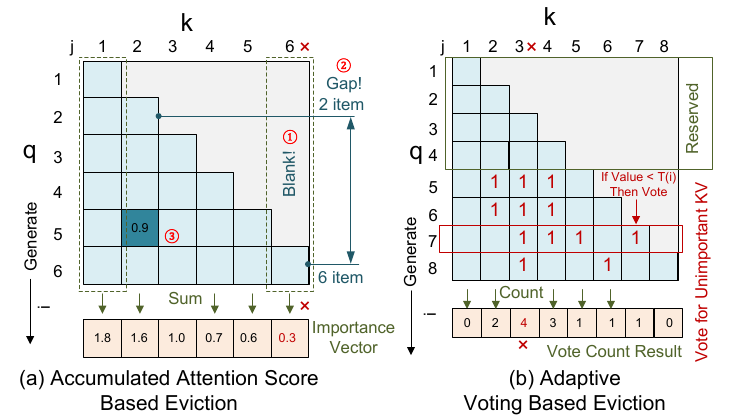}
\vspace{-15pt}
\caption{A comparison between the influential accumulation-based eviction and our voting-based strategy.}
\vspace{-10pt}
\label{Fig:comp}
\end{figure} 

\textbf{Limitations of Existing Method.} First, \underline{Item Count Bias}, as indicated by the red \ding{172} in the figure. The number of items summed across different columns varies significantly, with $6$ items for the first token, but only $1$ for the $6_{th}$ token due to causal attention, where $5$ elements are blank. This discrepancy leads to biased eviction, making recent KV tensors more likely to be pruned, which contradicts both intuition and studies suggesting that recent tokens are more important. Second, \underline{Criteria Bias}, denoted by \ding{173} in the figure. When examining different rows, $2$ items in the second row have a total attention score of $1$, resulting in a mean of $\frac{1}{2}$, while $6$ items in the $6_{th}$ row yield a mean of $\frac{1}{6}$. Hence, an attention score of $\frac{1}{3}$ may be considered insignificant in the second row but sufficiently important in the $6_{th}$. Due to these varying criteria, simply summing values across rows may not yield optimal results. Third, \underline{Outlier Bias}, indicated by \ding{174}. Some tokens have extremely high attention scores, indicating significance at certain positions but potentially less so at others. Nevertheless, the high scores are summed into the importance vector, leading to their consistent consideration as significant.

\textbf{Overview.} To address these challenges, we propose a \textit{voting-based} strategy to differentiate pivotal and unimportant kv vectors. An overview is shown in \figref{Fig:comp} (b). The method treats each generated token as a voter, fully leveraging the information in each attention score while adaptively adjusting the judgment criteria across positions. Specifically, alongside the attention score vector computation for each token, the algorithm also calculates an adaptive threshold for voting. If an attention score value is below this threshold, it is marked as unimportant, and a value of $1$ is set in the corresponding position of the vote count vector. This process updates the vote count vector continuously, and when the cache reaches its predefined size, the KV vector with the highest vote count is evicted. Additionally, a reserved length is maintained at the beginning, during which no voting occurs, aligning with the attention sink phenomenon \cite{xiao2023efficient}. 

Our evaluation shows that the proposed approach addresses the three above-mentioned limitations. First, it tends to preserve recent tokens; for example, in the step shown in the figure, position $7$ of KV is voted for at most twice, whereas position $3$ can be voted four times. Second, by applying an adaptive threshold relevant to each specific attention score vector, the criteria bias is effectively mitigated. Third, outlier bias is resolved by assigning a uniform weight of $1$ instead of relying on absolute values.
\begin{figure}[hbt]
\vspace{-20pt}
\centering
\includegraphics[width=88mm]{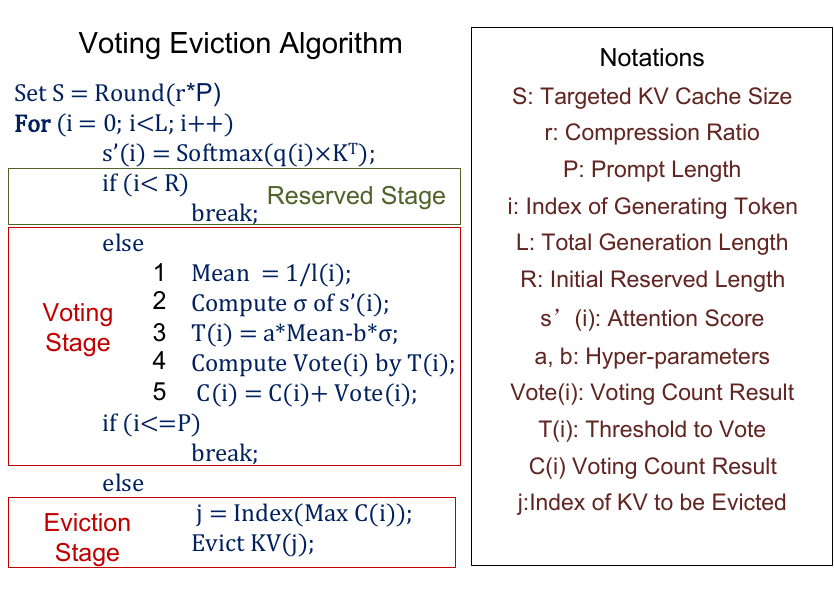}
\vspace{-15pt}
\caption{Voting-based KV cache eviction algorithm with notations on the right.}
\label{Fig:al}
\vspace{-10pt}
\end{figure} 

\textbf{Details.} The complete algorithm is illustrated in \figref{Fig:al}, with notations on the right, aligned with their order of occurrence. Common operations in LLM inference are omitted for clarity. The algorithm begins with an adjustable parameter $r$, provided by the user, representing the compression ratio, enabling a trade-off between aggressive compression and precision. The prompt length $P$ is determined once the prompt is input into the model. Based on $P$ and $r$, the target KV cache size $S$ is calculated. This configuration works well for tasks such as text summarization and question answering, though a fixed target size can also be set for language modeling as the sequence length extends from $1$ to the model's maximum capacity. 

The algorithm operates through both the prefilling phase and the generation phase. The \underline{prefilling phase} is divided into the reserved stage and the voting stage. Voting does not occur during the reserved stage, with its size $R$ set to $32$, establishing the lower bound for the KV cache. Line $3$ of the voting stage defines our adaptive threshold, as a linear combination of the mean and standard deviation of the attention scores. A sparse attention score results in a higher standard deviation, leading to a lower threshold, while a more evenly distributed score has a smaller standard deviation, producing a higher threshold—thus dynamically addressing the criteria bias. The hyper-parameters $a$ and $b$ can be fine-tuned through model-specific calibration without affecting deployment efficiency; values of $a=1$ and $b=0.2$ generally prove effective. Notably, the threshold may theoretically drop below zero, in which case the algorithm identifies the minimum attention score and votes accordingly. After processing the entire prompt, the algorithm transitions to the \underline{generation phase}, which includes voting and eviction stages. Voting continues for each current token, and the algorithm evicts the kv vector with the highest vote count. In cases where multiple positions have identical vote counts, the earliest position is selected for eviction. This strategy significantly accelerates the generation phase due to the reduced cache size.

\section{Flexible Dataflow Optimization}
\label{sec:dataflow}

\subsection{Flexible-product Dataflow}
\label{sec:product}
\textbf{Motivations.}
While dataflow optimization \cite{dao2022flashattention} and dataflow-optimized accelerator  \cite{wang2023cosa} for the LLM prefilling phase (or encoder-only Transformers) have been studied, a gap remains for the LLM generation phase. As mentioned in \secref{intro}, the wide range of dimension sizes, the dynamic inference process, and irregular memory access during transposition hinder efficiency. Notably, the current KV cache compression developed by the algorithm community significantly reduces storage overhead but is less effective in accelerating inference. Specifically, the GEMV of the attention process can be represented as $q \times K^T=s$ with the size $(1, d) \times (d, l) = (1, l)$ and $s' \times V=o$ with $(1, l) \times (l, d) = (1, d)$. With $l$ varying due to token growth during the inference process or KV cache reduction, existing architectures lack the efficiency to handle such variability. To address this, we return to mathematical interpretation for inspiration.
\begin{figure}[hbt]
\centering
\vspace{-10pt}
\includegraphics[width=88mm]{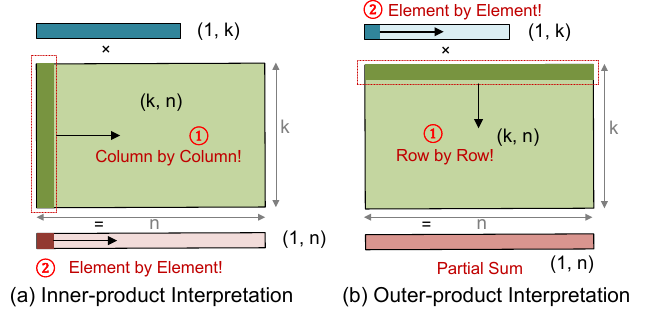}
\vspace{-15pt}
\caption{Two mathematical interpretations of GEMV.}
\label{Fig:math}
\vspace{-5pt}
\end{figure}
\begin{figure*}[hbt]
\centering
\vspace{-20pt}
\includegraphics[width=180mm]{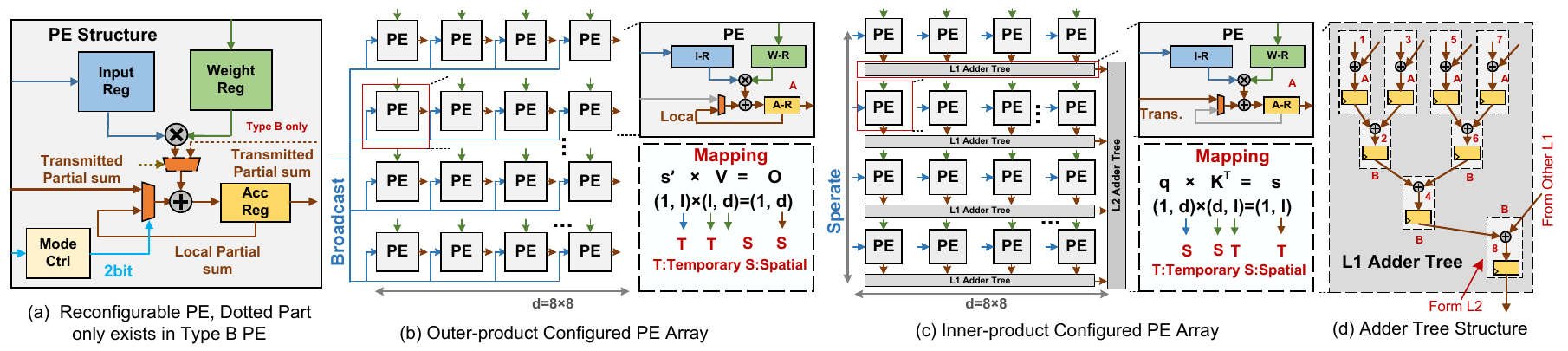}
\vspace{-10pt}
\caption{Runtime reconfigurable hardware architecture.}
\label{Fig:hardware}
\vspace{-5pt}
\end{figure*}
\textbf{Mathematical Definition:} There are two ways to interpret GEMV, as shown in \figref{Fig:math}. Given a general-purpose notation $(1, k) \times (k, n) = (1, n)$, \underline{for the inner-product method}, the entire input vector $(1, k)$ is multiplied by the weight matrix $(k, n)$, column by column, generating the output vector $(1, n)$ incrementally. This interpretation is widely used, especially in adder-tree-based implementations. \underline{For the outer-product method}, one element of the input vector is multiplied element-wise by each row of the weight matrix sequentially, generating a partial output vector $(1, n)$ of the same size as the final output. This partial output is accumulated into the result until all rows of the weight matrix are processed.
\begin{figure*}[ht]
\centering
\includegraphics[width=180mm]{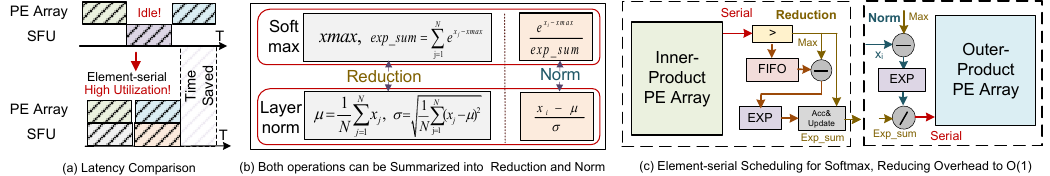}
\vspace{-5pt}
\caption{Element-serial scheduling.}
\label{Fig:es}
\vspace{-10pt}
\end{figure*}
\textbf{Flexible-product Inspiration:} \underline{For efficient memory access}, denoted by \ding{172} in \figref{Fig:math}, the weight memory access patterns of the two interpretations differ: one operates column-wise, while the other operates row-wise. Interestingly, it inspires us that storing the matrix uniformly, instead of performing transposition, a hybrid implementation on the \underline{computation engine} can have the equivalent effect to the transpose operation. Specifically, in the attention process, the KV matrix can be uniformly stored in $(l, d)$ format, where $l$ denotes the address, and $d$ elements share the same address. In each step, a $(1, d)$ vector of the K or V matrix can be accessed with a single memory operation, allowing the computation of $q \times K^T=s$ with  $(1, d) \times (d, 1)$ segment of $K^T$  per step, $l$ steps to realize $(1, d) \times (d, l) = (1, l)$ in an inner-product approach. Similarly, to compute $s' \times V=o$, element-wise multiplication is performed between $(1, 1)$ and $(1, d)$ segment of $V$ per step,  $l$ steps to realize $(1, l) \times (l, d) = (1, d)$, in an outer-product approach. This strategy avoids the costly transpose operation, ensuring a consistent storage format between the K and V matrices.

\underline{For efficient computation}, denoted by \ding{173} in \figref{Fig:math}, we observe that either the input or the output is accessed in an element-by-element manner, which offers flexibility in the corresponding dimension, indicating that arbitrary size in that dimension can be temporarily mapped to computation cycles. Specifically, for $q \times K^T=s$ with size $(1, d) \times (d, l) = (1, l)$, let $l$ correspond to $n$ and $d$ correspond to $k$ in \figref{Fig:math} (a); the variation in sequence length $l$ is effectively mapped to computation cycles. Similarly, for $s' \times V=o$ with $(1, l) \times (l, d) = (1, d)$, let $l$ and $d$ correspond to $k$ and $n$, respectively in \figref{Fig:math} (b), $l$ variations can be translated into computation cycles variation. In summary, analyzing the problem from both computational and storage perspectives led to a consistent conclusion: using the inner product for $q \times K^T$ and the outer product for $s' \times V$ is the optimal choice, leaving the primary challenge of designing a runtime-reconfigurable computation engine.

\textbf{Runtime Reconfigurable PE Array:}
\underline{Reconfigurable PE.} As illustrated in \figref{Fig:hardware} (a), each PE is governed by a 2-bit control signal, allowing it to perform local partial sum accumulation, transmit partial sums, clear registers, or disable. The multiplexers enable different modes to specify whether the partial sum is accumulated locally or received from other PEs. The dotted section, present only in type-B PEs, indicates that both adder inputs are sourced from other PEs. \underline{Outer-Product Configured PE Array.} PEs are structured as shown in \figref{Fig:hardware} (b) to perform an outer-product operation. Each PE operates independently with distinct weights, without spatial partial sum transmission, while sharing a broadcast input scalar. The PE array is set as $8 \times 8$, a typical finest dimension in modern LLMs. The structure and mapping of the PE are depicted on the right. For $s' \times V$, $l$ is temporally mapped to ensure flexibility, meaning that in each cycle, one scalar from $s'$ is broadcast to the PE array. The parameter $d$ is spatially assigned to each PE, with local accumulation of the partial sum. \underline{Inner-Product Configured PE Array.} In \figref{Fig:hardware} (c), weights and inputs are independently multiplied within each PE in parallel during each cycle, with partial sums transmitted across PEs. The PE adders are arranged into a two-level adder tree structure. The L1 adder tree receives multiplication results and performs accumulation at the row level, while the L2 adder tree aggregates the results from the L1 adder tree. For $q \times K^T$, $l$ is mapped spatially, allowing one output element of $s$ to be calculated per cycle, with the entire computation completed in $l$ cycles. The parameter $d$ is spatially mapped across the PE array. \underline{Hierarchical Adder Tree Implementation.} As shown in \figref{Fig:hardware} (d), the red annotations indicate the type and location of each PE in the row. The adder tree is configured with PEs $1, 3, 5, 7$ designated as type-A, where one input to the adder is the local multiplication result, and the second input is the multiplication output from PEs $2, 4, 6, 8$, respectively. PEs $2, 4, 6, 8$ are type-B, with both operands sourced from other PEs. The L2 adder tree follows similar design principles as the L1 adder trees, with the $8_{th}$ PE functioning as the local adder within a PE, compared to the L1 adder tree.

\subsection{Element-serial Scheduling}
\label{sec:nonlinear}
\textbf{Motivations.} Nonlinear operations such as softmax and layernorm impose significant data dependencies, as computation cannot proceed until all relevant elements are mathematically ready. These \underline{data dependencies} impede parallel processing, and the \underline{inherent latency} delays subsequent operations. As shown in \figref{Fig:es}, conventional scheduling causes the PE array to remain idle during the computation of softmax or layernorm in the Special Function Unit (SFU). Although increasing the computation units can reduce latency, it leads to higher overhead. Previous research \cite{9586134,yu2021nnlutneuralapproximationnonlinear} suggests using approximate units to mitigate costs. In contrast, we innovatively address the issue by co-designing dataflow scheduling and architecture, aiming to minimize the number of SFUs while eliminating latency, as illustrated in \figref{Fig:es} (a).
\begin{figure}[b]
\vspace{-15pt}
\centering
\includegraphics[width=88mm]{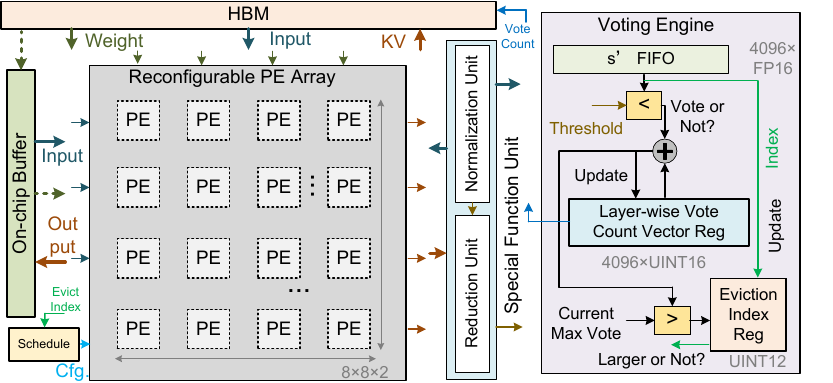}
\vspace{-10pt}
\caption{Hardware architecture of \textit{VEDA}, with blue, green, and red denote input weight and output.}
\label{Fig:arch}
\vspace{-20pt}
\end{figure}
\textbf{Element-serial Scheduling.} We summarize both softmax and layernorm operations into two stages: reduction and normalization (norm). In the reduction stage, elements are condensed into two numerical characteristics: maximum and exponent summation ($exp\_sum$) for softmax, and mean and standard deviation for layernorm. The normalization stage involves element-wise computations—subtraction, exponentiation, and division for softmax; subtraction and division for layernorm. Drawing from the inspiration of the element-by-element manner discussed in \secref{sec:product}, we propose to schedule the reduction stage element-by-element, running in parallel with PE array computations, which performs operations on the serial output of an inner-product-configured PE array; and scheduling the normalization on the serial input of an outer-product-configured PE array, concurrently with PE array computations as well. \textbf{A specific example} of softmax is demonstrated in \figref{Fig:es} (c). During the computation of \underline{$q \times K = s$} using the inner product in the PE array, the reduction unit in the SFU receives the serial, element-by-element output. It then identifies the maximum value for each tile, updates the maximum, and stores the tile in a FIFO. Once a tile is stored, its maximum is obtained, and this current maximum is used to update the $exp\_sum$, the process is similar to \cite{milakov2018online}. When the computation of all $l$ elements is completed, the final $exp\_sum$ is determined in just a few cycles. For \underline{$s' \times V = o$}, continuing with the inner product would costly require the concurrent preparation of $(1, l)$ softmax normalization results. Therefore, the outer product configuration facilitates element-by-element serial input into the PE array, enabling parallel processing in SFU. Notably, with a PE array of size $d$, a single SFU suffices to remove latency in \textit{element-serial scheduling} as shown in \figref{Fig:es} (a), and reduce the SFU cost from $O(N)$ to $O(1)$. The same technique applies to layernorm: inner-product implemented GEMV before reduction, the sum of element and the sum of element square is computed simultaneously through the element-serial output to obtain mean and variance; and outer-product implemented GEMV after normalization, the subtraction, and division are performed through element-serial input.  Since both softmax and layernorm are positioned between GEMV operations, they effectively define the optimal dataflow for the entire LLM: Please refer to \figref{Fig:LLM}, where green indicates the inner product and blue represents the outer product.

\section{Hardware Architecture}

Combining dataflow, hardware, and algorithm optimizations, we designed \textit{VEDA}, as depicted in \figref{Fig:arch}. The architecture comprises a PE array, Special Function Unit, voting engine, scheduler, and HBM serving as the off-chip memory. Detailed hardware specifications are provided in \tabref{tb:hardware}. \textit{VEDA} supports both prefilling and generation phases. \underline{For storage}, weights are fetched to the on-chip buffer and reused across tokens to minimize off-chip memory access for the prefilling phase; however, during the generation phase, they are directly input into the PE array. Thanks to the effective transposition elimination through flexible dataflow, the KV cache is uniformly stored in $(l, d)$, ensuring the high utilization of HBM bandwidth, they need to be stored off-chip, and they can be regarded as weight for the PE array during the generation phase. Additionally, the vote count also needs to be stored and accessed from off-chip to reduce the on-chip overhead. \underline{For computation}, the GEMM of the prefilling phase is achieved through multiple GEMVs, and weight is reused across GEMVs through buffering. Another significant benefit of GEMV implementation is that, compared with GEMM kernel computation, the effective operations of the attention process are reduced to half due to the causality, our flexible PE array can directly skip computing the upper triangle half of the attention score. 
\begin{figure*}[ht]
\vspace{-20pt}
\centering
\includegraphics[width=180mm]{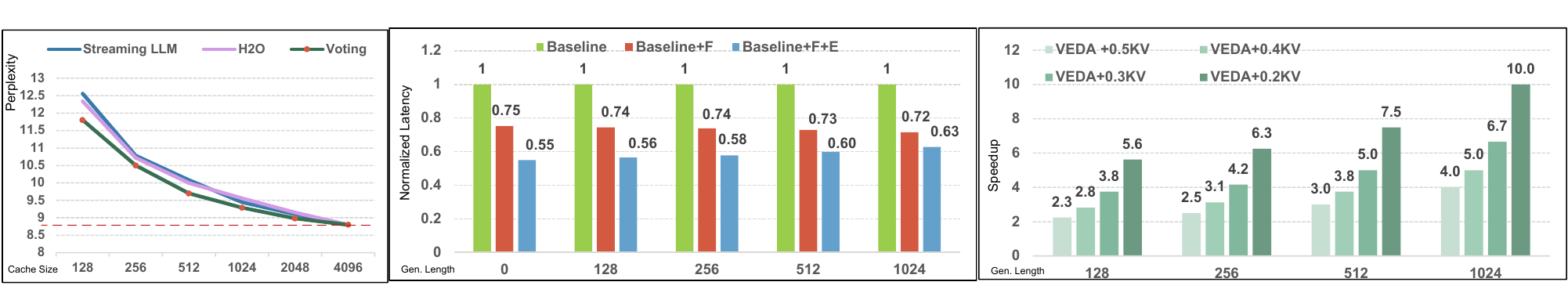}
\vspace{-10pt}
\caption{Experiments. Left: Language modeling evaluation. Center: Ablation study of dataflow optimization. Right: Speedup by voting-based cache eviction.}
\label{Fig:exp}
\vspace{-15pt}
\end{figure*}

Since the primary modules have been introduced, we will now focus on the \underline{voting engine's mechanism}. It receives the softmax result $s'$ (which also serves as input for $s'V$) and stores it in the FIFO. Simultaneously, $s'$ is internally input into the reduction unit to compute the mean and standard deviation. Once stored in the FIFO, the reduction unit also finishes the computation of mean and standard deviation, providing a threshold. Each element is then sequentially output from the FIFO to determine voting eligibility, update the previous vote counts, and adjust the registered vote count (hardware implemented by buffer) accordingly. During the prefilling phase, no eviction is performed. However, in the generation phase, the vote count is compared with the current maximum vote to update the eviction index. Voting operates layer-wise, meaning that all heads are aggregated and averaged. Once $s'$ across all heads is processed, the final eviction index is established, indicating that a particular address of kv vectors will no longer be accessed. The voting engine consistently operates in parallel, ensuring it does not affect the computations of other modules. 

\section{Evaluation}
\textbf{Experiment Setup.}
\underline{Hardware:} We utilize SystemVerilog to implement the \textit{VEDA} architecture, using FP16 as the default arithmetic format. The design is synthesized using the Design Compiler under the TSMC $28$nm CMOS library with 1GHz to evaluate area and power consumption. A $256$GB/s HBM is employed through Ramulator \cite{kim2015ramulator} to model the off-chip memory. For the on-chip buffer and FIFO, CACTI \cite{muralimanohar2009cacti} is used to estimate energy consumption and area.  A cycle-accurate performance model is employed to measure latency, which is cross-validated with RTL simulations. The hardware metrics are detailed in \tabref{tb:hardware}, demonstrating that PE and buffer dominate the area and power, SFU consumes less than $3\%$ due to \textit{element-serial scheduling}, and the voting engine incurs a small $6.5\%$ of overhead. \underline{Software:} We choose the influential Llama-2 $7$B model, with a maximum sequence length of $4096$, and use the language modeling task—a standard benchmark for assessing LLM performance—to validate our algorithm's feasibility. The primary evaluation metric is perplexity, with lower values indicating better performance.

\begin{table}[ht]
    \centering
    \vspace{-10pt}
    \caption{Hardware details, with area and power breakdown of VEDA}
    \label{tb:hardware}
       \scalebox{0.95}{
\begin{tabular}{c|ccc}
\hline
Modules        & Parameters                     & Area/$mm^2$      & Power/$mW$      \\ \hline
PE Array       & 8*8*2  Reconfigurable PEs      & 0.493        & 175.64        \\ \hline
\begin{tabular}[c]{@{}c@{}}Voting \\ Engine\end{tabular} &
  \begin{tabular}[c]{@{}c@{}}4096*16bit FIFO\\ 4096*16bit Vote Buffer\\ \& Others\end{tabular} &
  0.069 &
  26.41 \\ \hline
\begin{tabular}[c]{@{}c@{}}Special \\ Function Unit\end{tabular} &
  \begin{tabular}[c]{@{}c@{}}2 EXP, 2Divider, 1Sqrt\\ \& 2Multiplier and 4Adder\\ 32×16bit FIFO\end{tabular} &
  0.029 &
  13.19 \\ \hline
Schedule &
  \begin{tabular}[c]{@{}c@{}}System Control \& \\ PE Array Config\end{tabular} &
  0.041 &
  11.20 \\ \hline
On-chip Buffer & 256KB SRAM                     & 0.426        & 148.82        \\ \hline
Off-chip DRAM  & \multicolumn{3}{c}{HBM, Bandwidth=256GB/s}                    \\ \hline
Total          & \multicolumn{3}{c}{TSMC 28$nm$: Area=1.058$mm^2$  Power = 375.26$mW$} \\ \hline
\end{tabular}}
    \vspace{-5pt}
\end{table}

\textbf{Algorithm Evaluation:} We report perplexity results using 1000 samples from the PG-19 dataset \cite{rae2019compressive}, a widely-used benchmark for LLM language modeling evaluation. We compare our \textit{voting-based} cache eviction strategy with the notable Streaming LLM \cite{xiao2023efficient} and H2O \cite{zhang2023h2o} across various cache sizes. As depicted in \figref{Fig:exp} (left), the \textit{voting-based} eviction consistently outperforms both H2O and Streaming LLM, owing to the voting mechanism's ability to adaptively eliminate three types of bias and select the optimal key-value vector for eviction. Furthermore, even when the KV cache is reduced to $0.1$ of the original length ($4096$), the accuracy loss remains negligible, demonstrating the feasibility of our algorithm.

\textbf{Ablation Study.} 
\underline{For dataflow and reconfigurable hardware gain}, to assess the impact of our flexible architecture (F) and \textit{element-serial scheduling} (E), we use a conventional adder-tree-based architecture as the baseline, similar to $A^3$ \cite{a3} Transformer accelerator, in which the softmax operation functions as a pipeline stage with inherent latency. The conventional adder-tree-based design is constrained to a fixed dataflow for inner product computations, lacking flexibility. To ensure a fair comparison, accelerators are configured to have identical peak throughput and the same number of SFUs as \textit{VEDA}. We employ the Llama-2 $7$B model with a prompt length of $512$ for the prefilling phase and generate sequences ranging from $0$ to $1024$ tokens in the generation phase. The latency of the attention process is averaged over tokens during the generation phase, and depicted in \figref{Fig:exp} (center), the \textit{flexible-product} approach consistently achieves over a $25\%$ latency reduction due to higher utilization across varying lengths. The \textit{element-serial scheduling} further reduces latency to approximately $60\%$. \underline{For algorithmic gain}, we use the same $512$-length input and vary the maximum generated sequence length between $128$ and $1024$, averaging the latency of the attention process over tokens. The baseline in this evaluation is \textit{VEDA} without cache eviction. The \textit{voting-based} approach effectively maintains a constant KV length ($512 \times \text{ratio}$) throughout the generation phase, in contrast, the baseline's KV cache grows with each step. This approach achieves a $2.3\sim10.0\times$ speedup over the baseline.

\textbf{Comparison.} \tabref{tb:comp} shows that, compared with SOTA accelerators, thanks to the high utilization powered by flexible dataflow, and low-cost SFU and voting, \textit{VEDA} attains the smallest area cost, and the highest energy efficiency (it remains true after technology scaling \cite{sarangi2021deepscaletool}). Compared with the SOTA edge GPU, \textit{VEDA} achieves $38.8\times$ energy efficiency. One \textit{VEDA} has a throughput of $18.6$ tokens/s. Compensating for the resource discrepancy, $8-VEDA$ will have a $2.86\times$ throughput improvement over GPU. The evaluations demonstrate the feasibility of LLMs edge deployment through \textit{VEDA}.

\begin{table}[ht]
\vspace{-10pt}
    \centering
    \caption{Comparison with related accelerators}
    \label{tb:comp}
       \scalebox{0.95}{
\begin{tabular}{cccc}
\hline
\rowcolor[HTML]{EFEFEF} 
Accelerator Comparison            & Sanger        & Spatten         & \textit{VEDA}             \\ \hline
Support                        & Attention     & Transformer     & \textbf{LLM}     \\
Technology{[}$nm${]}                & 55          & 40            & \textbf{28}    \\
Area{[}$mm^2${]}                      & 16.9          & 1.55            & \textbf{1.06}    \\
Throughput{[}$GOPS${]}              & 529           & 360             & \textbf{245}     \\
Energy Efficiency{[}$GOPS/W${]}     & 192           & 382             & \textbf{653}     \\ \hline
\rowcolor[HTML]{EFEFEF} 
\multicolumn{4}{c}{\cellcolor[HTML]{EFEFEF}End-to-end Comparison with NVIDIA 4090 GPU} \\ \hline
\multicolumn{4}{c}{Average Energy Efficiency:  38.8× (Core+Off-chip HBM)}                       \\
\multicolumn{4}{c}{Average Throughput:  1-\textit{VEDA}: 18.6 tokens/s,  8-\textit{VEDA}:2.86× over GPU}                     \\ \hline
\end{tabular}
}
\vspace{-10pt}
\end{table}
\section{Conclusion}
In conclusion, this work addresses the computational challenges of LLM inference on edge devices through  \textit{voting-based} KV cache eviction, \textit{flexible-product} dataflow, \textit{element-serial scheduling}, a pioneering algorithm-hardware-dataflow tri-optimizations. 
The resulting accelerator, \textit{VEDA}, demonstrates significant performance improvements, enabling efficient LLM inference on resource-constrained platforms, and promoting real-time processing and data privacy.
\newpage
\bibliographystyle{IEEEtranS}
\bibliography{refs}
\end{document}